\documentclass[preprint]{aastex631}
\usepackage{url}
\def\UrlBreaks{\do\A\do\B\do\C\do\D\do\E\do\F\do\G\do\H\do\I\do\J
\do\K\do\L\do\M\do\N\do\O\do\P\do\Q\do\R\do\S\do\T\do\U\do\V
\do\W\do\X\do\Y\do\Z\do\[\do\\\do\]\do\^\do\_\do\`\do\a\do\b
\do\c\do\d\do\e\do\f\do\g\do\h\do\i\do\j\do\k\do\l\do\m\do\n
\do\o\do\p\do\q\do\r\do\s\do\t\do\u\do\v\do\w\do\x\do\y\do\z
\do\.\do\@\do\\\do\/\do\!\do\_\do\|\do\;\do\>\do\]\do\)\do\,
\do\?\do\'\do+\do\=\do\#}
\shortauthors{Zheng et al.}
\begin{document}

\title{Twin extreme ultraviolet waves in the solar corona}
\author{Ruisheng Zheng}
\affiliation{Shandong Key Laboratory of Optical Astronomy and Solar-Terrestrial Environment, School of Space Science and Physics, Institute of Space Sciences, Shandong University, Weihai, Shandong, 264209, China}

\author{Bing Wang}
\affiliation{Shandong Key Laboratory of Optical Astronomy and Solar-Terrestrial Environment, School of Space Science and Physics, Institute of Space Sciences, Shandong University, Weihai, Shandong, 264209, China}

\author{Liang Zhang}
\affiliation{Shandong Key Laboratory of Optical Astronomy and Solar-Terrestrial Environment, School of Space Science and Physics, Institute of Space Sciences, Shandong University, Weihai, Shandong, 264209, China}

\author{Yao Chen}
\affiliation{Shandong Key Laboratory of Optical Astronomy and Solar-Terrestrial Environment, School of Space Science and Physics, Institute of Space Sciences, Shandong University, Weihai, Shandong, 264209, China}

\author{Robertus Erd\'{e}lyi}
\affil{Solar Physics \& Space Plasma Research Center (SP2RC), School of Mathematics and Statistics, University of Sheffield, Hicks Building, Hounsfield Road, S3 7RH, United Kingdom}
\affil{Department of Astronomy, E\"otv\"os Lor\'and University, 1/A P{\'a}zm{\'a}ny P{\'e}ter s{\'e}t{\'a}ny, H-1117 Budapest, Hungary}
\affil{Gyula Bay Zolt\'an Solar Observatory (GSO), Hungarian Solar Physics Foundation (HSPF), Pet\H{o}fi t\'er 3, H-5700 Gyula, Hungary}

\correspondingauthor{Ruisheng Zheng; Robertus Erd\'{e}lyi}
\email{ruishengzheng@sdu.edu.cn; r.von.fay-siebenburgen@sheffield.ac.uk}

\begin{abstract}
Solar extreme ultraviolet (EUV) waves are spectacular propagating disturbances with EUV enhancements in annular shapes in the solar corona. These EUV waves carry critical information about the coronal magnetised plasma that can shed light on the elusive physical parameters (e.g. the magnetic field strength) by global solar coronal magneto-seismology. EUV waves are closely associated with a wide range of solar atmospheric eruptions, from violent flares and coronal mass ejections (CMEs) to less energetic plasma jets or mini-filament eruptions. However, the physical nature and driving mechanism of EUV waves is still controversial. Here, we report the unique discovery of twin EUV waves (TEWs) that were formed in a single eruption with observations from two different perspectives. In all earlier studies, a single eruption was associated at most with a single EUV wave. The newly found TEWs urge to re-visit our theoretical understanding about the underlying formation mechanism(s) of coronal EUV waves. Two distinct scenarios of TEWs were found. In the first scenario, the two waves were separately associated with a filament eruption and a precursor jet, while in another scenario the two waves were successively associated with a filament eruption. Hence, we label these distinguished scenarios as "fraternal TEWs" and "identical TEWs", respectively. Further, we also suggest that impulsive lateral expansions of two distinct groups of coronal loops are critical to the formation of TEWs in a single eruption.
\end{abstract}

\keywords{Sun: activity --- Sun: corona --- Sun: coronal mass ejections (CMEs)}

\section{Introduction}
The solar corona is filled with dynamic magnetized plasma, and hosts a variety of disturbances/waves that can be driven by all motions of the ubiquitous plasma. One of the most spectacular waves is the extreme ultraviolet (EUV) wave that appears as the traveling coronal disturbance from its erupting source region in the solar atmosphere. Hints about the existence of EUV waves were initially deduced from indirect evidences of sympathetic solar activities, type II radio bursts, and H$\alpha$ Moreton waves. Later, EUV waves were finally detected directly with the Extreme Ultraviolet Imaging Telescope (EIT; Delaboudini{\`e}re et al. 1995) on-board the SOHO spacecraft (Moses et al. 1997; Thompson et al. 1998; Warmuth 2015) in 1997. EUV waves can provide potential diagnostics of the coronal magnetic field strengths and can be used to estimate coronal plasma parameters that are hard to observe directly (Ballai 2007; Kwon et al. 2013). 

Although there is ongoing debate about the physical nature of these waves, an EUV wave is always interpreted either as a true wave or as a pseudo wave. The interpretations of true waves include linear/nonlinear fast-mode magnetohydrodynamic (MHD) waves, slow-mode waves, and soliton-like waves (Ofman \& Thompson 2002; Wills-Davey et al. 2007, Wang et al. 2009). On the other hand, the methods of employing pseudo waves embrace e.g. the magnetic field-line stretching, Joule heating in current shells, and continuous small-scale reconnections (Chen et al. 2002; Delaboudini{\`e}re et al. 2008; Attrill et al. 2007). Benefiting from high-quality observations from the Solar Terrestrial Relations Observatory(STEREO; Kaiser et al. 2008) and the Solar Dynamics Observatory (SDO; Pesnell et al. 2012), EUV waves have been best interpreted as a bimodal composition of an outer fast-mode MHD wave and an inner non-wave component of coronal mass ejections (CMEs) in a hybrid model (Liu et al. 2010, Chen \& Wu 2011, Downs et al. 2012, Liu \& Ofman 2014, Mei et al. 2020). More details about the nature of EUV waves are in recent reviews (Gallagher \& Long 2011, Patsourakos \& Vourlidas 2012; Liu \& Ofman 2014, Warmuth 2015, Chen 2016, Long et al. 2017, Shen et al. 2020). It is widely recognized by now that the EUV waves are associated with a variety of energetic eruptions (e.g. CMEs and flares), and small-scale EUV waves are closely associated with small-scale ejections (e.g. jets and eruptions of mini-filaments) (Zheng et al. 2012a, b). In addition, it is suggested that the formation of an EUV wave strongly depends on the rapid expansion of overlying coronal loops ahead of an erupting core (Zheng et al. 2019, 2020).

To date, it has only been observed that each single eruption triggered only a single EUV wave. In this study, we now provide new observational evidence for two scenarios of TEWs in a single eruption. TEWs in the first scenario were separately associated with a filament eruption and its precursor jet, while those in the second scenario were successively related to another filament eruption. Hence, we refer to these cases as "fraternal TEWs" and "identical TEWs", respectively. In the following text, the term "EUV wave" is simply abbreviated as ``wave".

\section{Observations}
For the first scenario, the filament eruption occurred beyond the northwest limb on 2010 August 18, and involved two EUV waves, a C4.5 flare, and a partial halo CME ({\url{https://cdaw.gsfc.nasa.gov/CME\_list/UNIVERSAL/2010\_08/univ2010\_08.html}}) with a linear speed of 1471 km s$^{-1}$. The source region, located at the mixture polarities of National Oceanic and Atmospheric Administration (NOAA) Active Region (AR) 11093 and 11099, was confirmed by the ARs on the solar disk on August 14 in Helioseismic and Magnetic Imager (Scherrer et al. 2012) magnetograms.

For the second scenario, the filament eruption occurred at AR 11228 near the northeast limb on 2011 June 1. The eruption involved two EUV waves, and a C2.6 flare. During the eruption, there was a slow CME (\url{https://cdaw.gsfc.nasa.gov/CME\_list/UNIVERSAL/2011\_06/univ2011\_06.html}) with a linear speed of 259 km s$^{-1}$, which was better seen in the view of COR1 (\url{https://cor1.gsfc.nasa.gov/catalog}).  However, the CME originated from another eruption in AR 11227 in the south hemisphere.

Besides the view provided by SDO, the two scenarios were also captured by the spacecrafts A and B of STEREO, respectively. The positions of STEREOs (A and B) and SDO (Earth) for two cases are shown in Figure 1. We used the EUV observations from the Atmospheric Imaging Assembly (AIA; Lemen et al. 2012) on-board SDO and from the Extreme Ultraviolet Imager (EUVI; Howard et al. 2008) on the STEREO. The AIA instrument contains seven EUV wavelengths that involve a wide range of temperature coverage, and its images have a field of view (FOV) of 1.5 $R_\odot$ with a pixel resolution of 0.6$"$ and a cadence of 12 s. The EUVI images have a pixel resolution of 1.58$"$, and their cadences are 2.5 minutes for both the 195~{\AA} and 304~{\AA}. In addition, the filaments and the associated jet, observed in the first scenario, were also recorded by the H$\alpha$ filtergrams from the Solar Magnetic Activity Research Telescope (SMART; UeNo et al. 2004) with a solar imaging system of Solar Dynamics Doppler Imager (Ichimoto et al. 2017) at Hida Observatory. The H$\alpha$ images have a cadence of 1 minute and a pixel size of $\sim$1$"$.

To display better the faint wavefronts in the first scenario, the EUV images are processed by the intensity-normalization method, where the value at a pixel is normalized by all the intensities at this same pixel among a set of the original images. For the diffuse wavefronts in the second scenario, the EUV images are subtracted by a fixed image and the previous image with a fixed time gap. The evolutions of the EUV waves and the associated erupting cores and loops, are analysed by the time-slice approach (Liu et al. 2010), in which a time-distance plot was constructed with a stack of slices along a selected direction for a set of images. 8 slices (S0-S8) are chosen,  and their start points are indicated by triangles. In the first scenario, S0 is in the original ejection direction of the jet, and S1 is in the vertical northward direction, and S2 is an arc at the height of 0.1 $R_\odot$ over the northwest limb, and S3 is in the lateral expanding direction of the north end of L1, and S4 is an arc path with the height of 0.1 $R_\odot$ over the southwest limb, and S5 is in the direction of $20^\circ$ counted clockwise from the south in the perspective of EUVI-A. In the second scenario, S6 is in the direction of the erupting filament, and S7-S8 are in the vertical southward directions in the view of AIA and EUVI-B, respectively. The speeds are obtained by fitting with the linear ({\it linfit.pro}) function, assuming a measurement uncertainty of 4 pixels ($\sim1.74$ Mm) for the selected points. On the other hand, the coronal magnetic filed lines are also extrapolated with {\it pfss\_trace\_field.pro} and {\it pfss\_draw\_field.pro}, parts of the PFSS package of SolarSoftWare (see the template routine of {\it pfss\_sample1.pro}).

\section{Results}
\subsection{Fraternal TEWs}
The fraternal TEWs reported here occurred on 2010 August 18, and the pre-eruption evolution is shown in SMART H$\alpha$, AIA 304~{\AA} and EUVI-A 304~{\AA} (Figure 2). Half an hour prior to the eruption ($\sim$04:41 UT), two small filaments (white arrows in panels (a)-(b)) appeared suddenly and rose slowly from the source region behind the northwest limb. The two small filaments were clearly disconnected with different chiralities, one hour earlier in the disk view of EUVI-A (white arrows and black dotted lines in panel (c)). Interestingly, a jet (marked by cyan arrows) and a new long filament (green arrows) occurred during the filament ascent, and the jet clearly originated from the junction (see the blue arrow) of two small filaments in the disk view of EUVI-A (panels (d)-(f)). It is likely that the jet and the newly-formed long filament resulted from the collision and subsequent magnetic reconnection of the two small filaments. Immediately, this newly-formed long filament lifted southwestwards in a non-radial direction (green arrows) (panels (g)-(h)). On the other hand, the jet moved northwards (cyan arrows in bottom panels), and had an initial speed of $\sim$207 km s$^{-1}$ (panel (h1)) along S0 (panel (h)). The measured jet speed is highly supersonic and is consistent with the Alfv\'en speeds (an estimated speed of $\sim$210 km s$^{-1}$, assuming the electron number density of $10^{8} cm^{-3}$ and the radial coronal magnetic field component of $1 G$ at a height of $0.06 R_\odot$).

Ahead of the ejecting jet, a faint wave (W1) formed and is visualised in intensity-normalized images of AIA 211~{\AA} and EUVI-A 195~{\AA} (Figure 3 and Movie S1). W1 mainly propagated northeastwards in the lower corona, and was much clearer in the disk view of EUVI-A, due to the projection effect (red arrows in panels (a)-(b)). There is a loop-like dimming structure (the yellow arrow) propagating visualised in difference images of AIA 193~{\AA} (panel (c)), what is an indicator of the expansion of a group of overlying coronal loops (L1). On the disk view, the dimmings around the north end of L1 follows the faint wavefront (white and red arrows in panel (d)), which reflects a close relation between the propagating W1 and the expanding L1. In the time-distance plot (panel (e)) along S1, the expansion speed of L1 is estimated to be a supersonic $\sim$167 km s$^{-1}$ (dotted lines with asterisks), and there is also a close temporal and spacial relationship between the jet and the onset of the L1 expansion (the blue asterisk and the green arrow). In the time-distance plots of intensity-normalized images along S2 and S3 (panels (f)-(g)), it is shown that W1 started at $\sim$04:58 UT (blue asterisks) with a nearly constant speed of $\sim$230 km s$^{-1}$, and the wavefront in EUVI-A 195~{\AA} is stronger than its counterpart in AIA 211~{\AA}.

In the meantime of the northwards propagation of W1, the newly-formed long filament also kept rising and eventually erupted (Figure 4). The filament (green arrows) lifted southwestwards and quickly vanished in AIA 171~{\AA} (panels (a)-(d)). However, a hot channel appeared in AIA 131 and 94~{\AA} (white arrows in panels (e)-(f)), which is an indicator of a high-temperature flux rope (Zhang et al. 2012). Meanwhile, there formed the current sheet containing plasma blobs beneath the flux rope (black arrows in panels (d)-(f)), which was studied in details by Takasao et al. (2012). The legs of the erupting flux rope (white arrows) remain still anchored in the solar surface, and the post-eruption cusp structure (cyan arrows) appeared in the eruption center (panels (h)-(i)).

Surprisingly, the flux rope eruption was followed by another wave (W2) that mainly travelled southwards (see Figure 5 and Movie S2). W2 was much stronger than W1, which is apparent in intensity-normalized images in AIA 211~{\AA} and EUVI-A 195~{\AA}. W2 was closely associated with the expansion of another group of coronal loops (L2; cyan arrows in top panels). W2 propagated southwards along the limb, as seen in the AIA view, and displayed an arc shape in the disk view (yellow and red arrows in panels (c)-(d)). In the time-distance plots along S4 and S5 (panels (e)-(f)), W2 shows a nearly constant velocity of $\sim$390 km s$^{-1}$ (yellow and red lines with asterisks) that is highly supersonic and likely super-Alfv\'enic, and set off at $\sim$05:20 UT (blue asterisks), closely following the flux rope eruption.

\subsection{Identical TEWs}
The identical TEWs occurred on 2011 June 1, and the related eruption is shown in Figure 6. The eruption region (the dotted box) is comprised of a central parasitic negative polarity (N1), and surrounding positive polarities that formed a circular neutral line, and a sunspot with negative polarity (N2) located at the west boundary (panel (a)). It is apparent that the filament consisted of two branches (white arrows) that lay along the circular neutral line, and are rooted at the central N1 and the strong positive polarity (P1) in the north (panels (b)-(c)). Likely due to some disturbances from P1, the filament erupted as a jet (red arrows in panels (d)-(f)). It is noteworthy that the escaping jet was guided by a bundle of coronal loops (L3; the yellow arrow) that connected the P1 and the N2, which was best seen in AIA 94~{\AA} (panel (d)). After the eruption, there appeared a cusp structure (the orange arrow) connecting P1 and the remote N2 over the flare loops (the green arrow), as a result of the explosion of the overlying L3 (panel (g)). Interestingly, the eruption failed and left a bunch of filament threads rooted at the P1 (red arrows in panels (h)-(i)), which indicates the confinement from a higher group of coronal loops.

Following the eruption, the taller group of confined loops (L4; dashed curves) is apparent in the running difference images, and interestingly, two waves successively formed in a short period (see Figure 7 and Movie S3). The first wave (W3) emerged from the expansion direction of L3, and the initially narrow wavefront then became diffuse (red arrows in panels (a)-(d)). Following the expansion of L4, the second wave (W4) formed at the south flank of the expanding L4,  $\sim$150 Mm behind W3 (the red and cyan arrows in panel (d)). The front of W4 developed into an arc shape some minutes later (the cyan arrow in panel (f)). Note that a strong wave (yellow arrows) came from the south, and interacted with W3-W4 propagating southwards (panels (e)-(f)). From the intensity-normalized time-distance plot in AIA~193{\AA} along S6 (panel (g)), the erupting filament speed of $\sim$250 km s$^{-1}$, again a highly supersonic value, is estimated after extricating from the L3 confinement, what then quickly began to fall back (the pink arrow) at $\sim$02:50 UT (the blue vertical line), likely due to the restriction by L4. In the running-difference time-distance plot in AIA~193{\AA} along S7 (panel (h)), W3 and W4 started at $\sim$02:44 and $\sim$02:49 UT (blue asterisks), respectively, and the likely both supersonic and super-Alfv\'enic speed of W3 ($\sim$495 km s$^{-1}$) was higher than that ($\sim$370 km s$^{-1}$) of W4, and two waves encountered the south strong wave (the yellow arrow). Due to the lower cadence, the wavefronts of W3-W4 consisted of some discrete brightenings (red and cyan arrows) in the running-difference time-distance plot along S8 (panel (i)).

\section{Conclusions and Discussion}
The observational results presented above are evidences for two distinct scenarios of TEWs that successively occurred in a single eruption in a short interval ($\sim$22 minutes and $\sim$5 minutes) and TEWs were confirmed by two different viewpoints from SDO and STEREO. W1-W4 had a linear speed in the range of $\sim$230-500 km s$^{-1}$ that is highly supersonic and likely super-Alfv\'enic giving away hints about their nature, which is then consistent with the interpretation of fast-mode MHD waves. Two scenarios show two different formation situations of these TEWs. In the first scenario, W1 and W2 were separately associated with two erupting portions (the precursor jet and the first filament eruption) that simultaneously formed during the magnetic reconnection process between two small filaments. In the second scenario, both W3 and W4 were associated with the same filament eruption in the form of blowout jet (Moore et al. 2010, Li et al. 2018). Therefore, we refer two scenarios as "fraternal TEWs" and "identical TEWs", respectively.

How did these TEWs formed? Firstly, we superimposed the magnetic field lines extrapolated by the potential field source surface (PFSS;  Schrijver \& De Rosa 2003) model on the intensity-normalized images for the fraternal TEWs, and, on difference images for the identical TEWs (see Figure 8). For the fraternal TEWs (top panels), W1 and W2 (yellow and red arrows) appear at the flanks of the yellow and red loops at different times. For the identical TEWs (bottom panels), the loop-like dimmings (the white arrow) and remote dimmings (purple arrows) probably indicate the explosion of the orange loops and the expansion of cyan loops. Hence, the orange, red, and orange, and cyan extrapolated loops are corresponding to the observations of L1, L2, L3, and L4, respectively.

Secondly, we proposed a generation sketch of the two distinct scenarios of TEWs (Figure 9), in which the selected important extrapolated coronal loops (yellow, red, orange, and cyan lines) are superimposed on the extrapolated magnetogram (top panels) and the HMI magnetogram (bottom panels).

For the fraternal TEWs (top panels), two small filaments (brown and pink lines) slowly rose and interacted with each other (the yellow star symbol), which simultaneously produced a jet and a long filament. The powerful jet moved northwards (the dashed arrow) and punched the northern overlying yellow L1, and the rapid lateral expansion of L1 gave birth of W1 (the cyan shade). On the other hand, the newly-formed long filament (the purple rope) erupted and forced the southern overlying red L2 to expand suddenly, and the W2 formed (the blue shade) at the east flank of the expanding green loops.

For the identical TEWs (bottom panels), the filament eruption initially pushed the lower overlying orange L3, and W3 formed (the yellow shade). After the explosion of L3, the erupting filament hit the higher overlying cyan L4, and W4 was formed (the green shade) at the south flank of the expanding L4. Therefore, we suggested that the four waves were driven by the rapid expansions of overlying loops (Figure 10 and 11 in Appendices).

TEWs in a single eruption are two fast-mode MHD waves, which is different from the two components of the fast-mode wave and the coronal reconfiguration signature in the hybrid model, see e.g. Liu et al. 2014, Long et al. 2017. On the other hand, there exist reports on multiple-wave phenomena, e.g. the quasi-periodic fast-propagating (QFP) waves and the homologous waves. QFP waves propagate as wave trains at the local Alfv\'{e}n speed along open field lines, and it is believed that QFP waves are excited by repetitive flaring energy releases (Liu et al. 2011, 2012). Meanwhile, homologous waves occur successively from the same place in a few hours or days, but they are generated by a series of distinct eruptions from the same region (Kienreich et al. 2011, Zheng et al. 2012b). In this study, we report TEWs triggered by the rapid expansion of overlying loops, and both waves were closely associated with a single eruption. Hence, TEWs are also intrinsically different from the QFP waves and homologous waves.

Why are TEWs rare, though they have the same physical nature and driving mechanism with that of a single wave? It is likely because the TEWs easily mixed together. Assuming that two groups of coronal loops stay near or expand in close directions, the two waves generated will mix and become hardly distinguishable from each other. If a wave-related strong eruption invokes a series of groups of coronal loops, then a series of wave segments will compose a large arc-like or circular wavefront. Hence, we further suggest that some visible wavefront consist of a series of undistinguishable wave segments that are triggered by a series of groups of coronal loops.

The discovery of TEWs likely indicate that more TEWs can be detected in future. We point out that the key for the formation of TEWs is the expansion that is divided in the directions of two separate groups of coronal loops in a single eruption. Actually, it is ubiquitous for the plasma motions in the solar corona that consists of building blocks of coronal loops. Hence, the EUV wave should be prevalent in the solar corona. However, the number of detectable EUV waves is far less than the expected one, which is possibly because most weak and small EUV waves are submerged by ubiquitous coronal loops around, or due to some waves may be at lower temperatures than what the current filters can detect.

\begin{acknowledgments}
SDO is a mission of NASA's Living With a Star Program. We gratefully acknowledge the usage of data from the SDO, STEREO, and from the ground-based SMART project. This work is supported by grants of NSFC 11790303 and 12073016.
\end{acknowledgments}

\clearpage

\begin{figure}
\epsscale{0.95}
\plotone{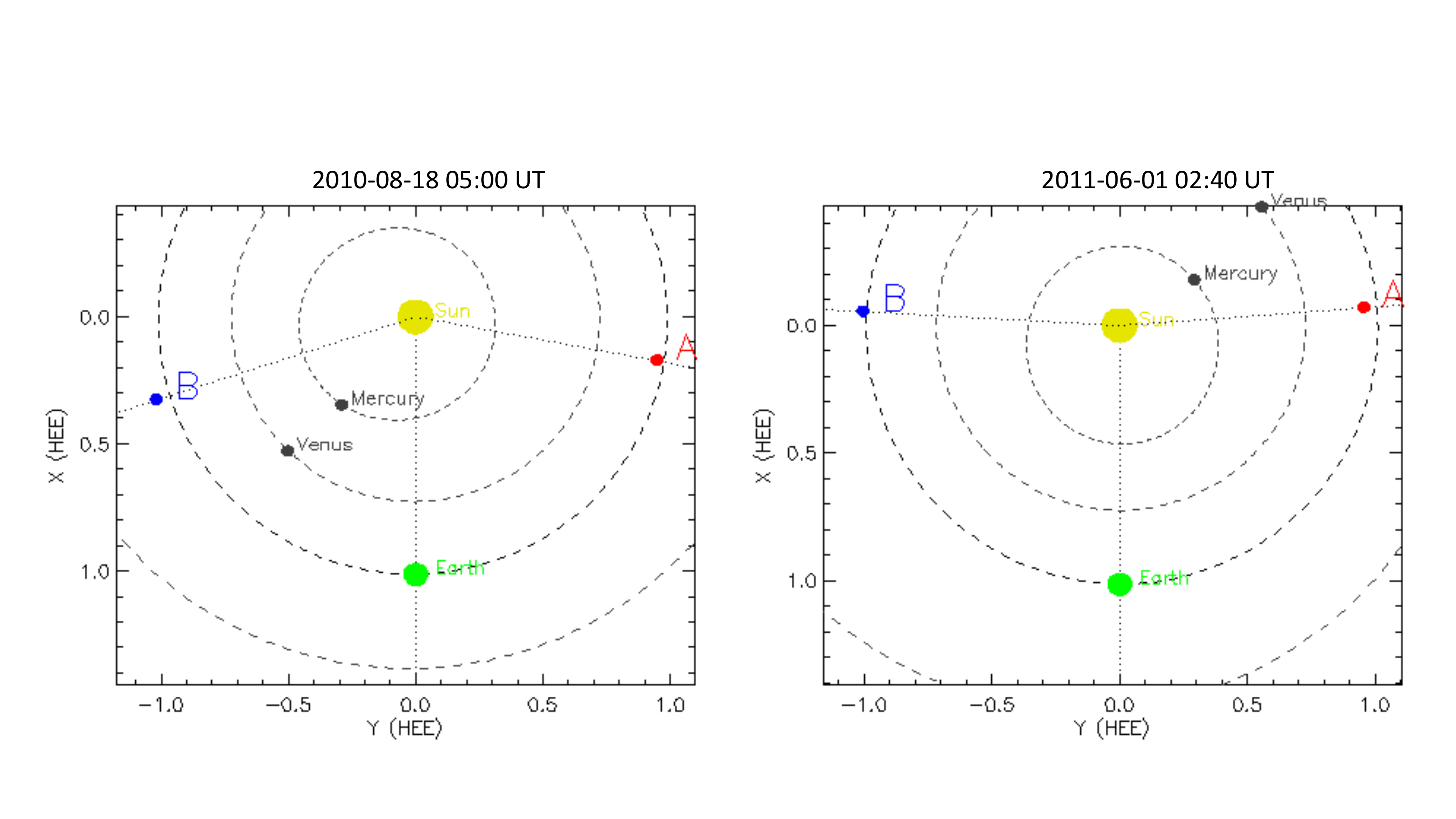}
\caption{Positions of SDO (Earth) and STEREOs (A and B) for two events ($https://stereo-ssc.nascom.nasa.gov/cgi-bin/make\_where\_gif$).
\label{f1}}
\end{figure}
\clearpage   

\begin{figure}
\epsscale{0.95} \plotone{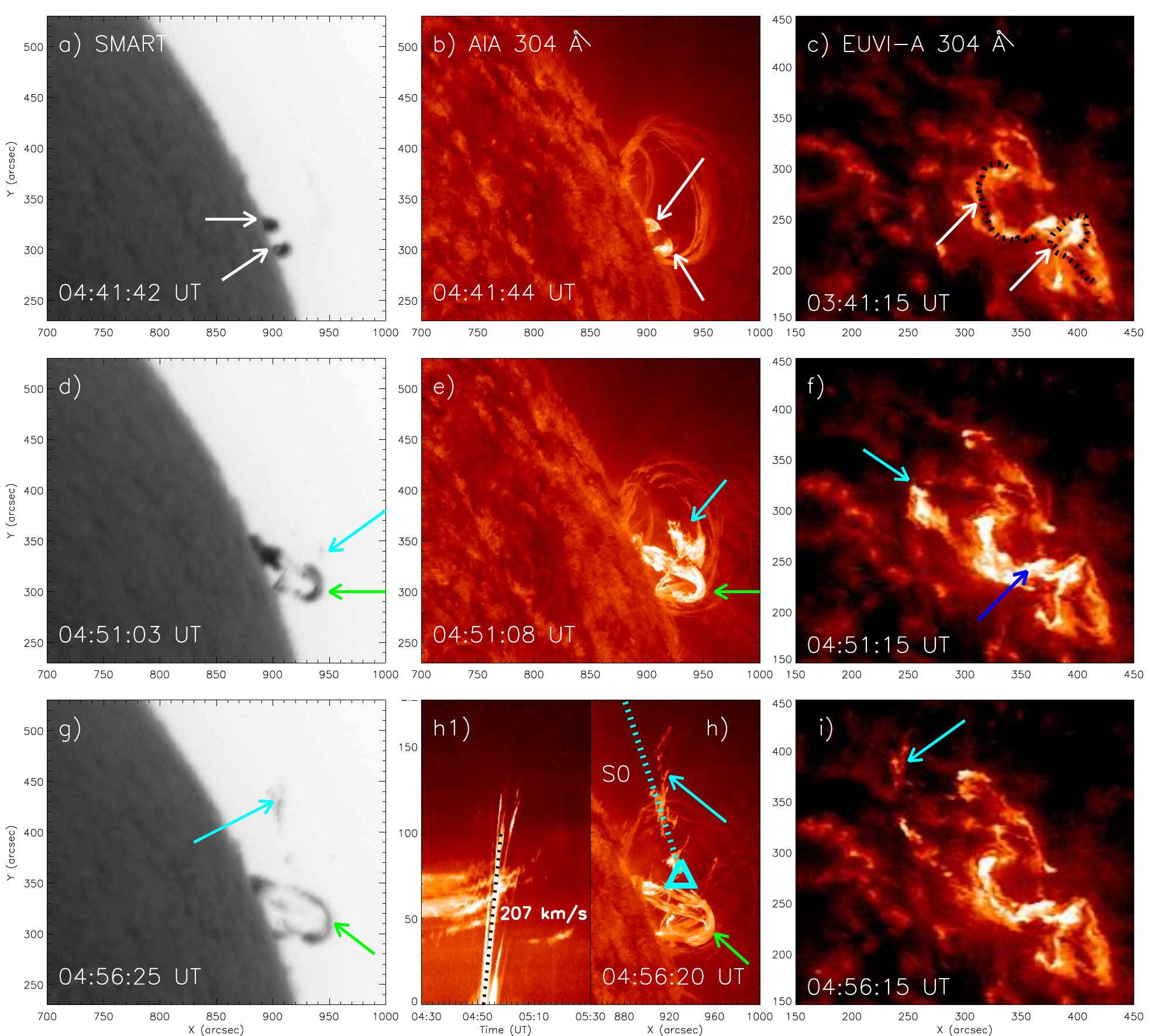}
\caption{The evolution of the jet in negative filtergrams of SMART H$\alpha$ (left panels) and in images of AIA 304~{\AA} (middle panels) and EUVI-A 304~{\AA} (right panels). Two small filaments and their junction are pointed out by white and blue arrows, respectively. The cyan arrows indicate the jet, and green arrows indicate the newly-formed long filament. The dotted line (S0) starting at a triangle is used to derive the jet speed that is shown in the time-distance plot of panel (h1).
\label{f2}}
\end{figure}

\clearpage

\begin{figure}
\epsscale{0.8}
\plotone{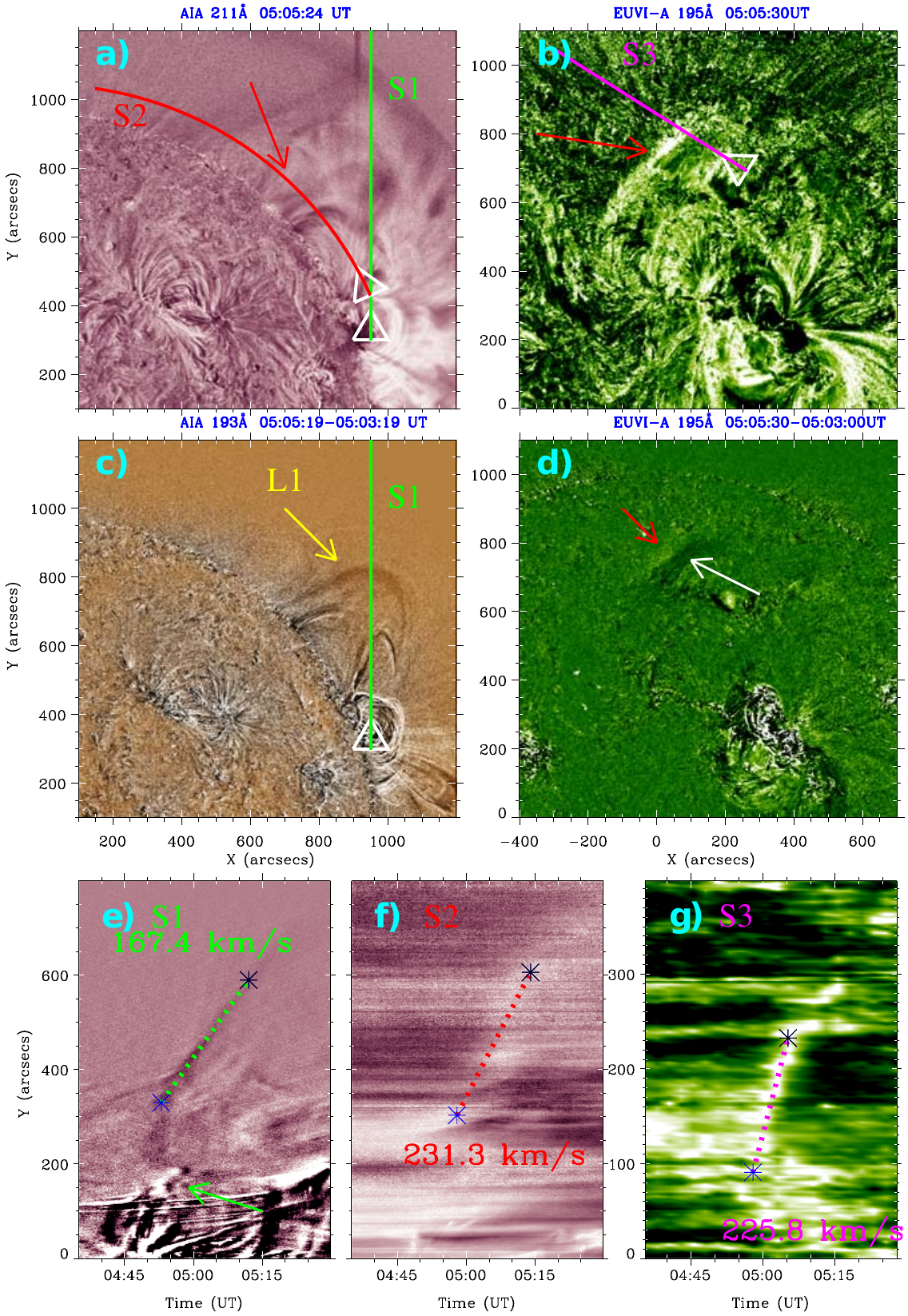}
\caption{W1 in intensity-normalized images between 04:30 and 06:30 UT in AIA 211 and 193~{\AA} (left panels) and EUVI-A 195~{\AA} (right panels). The green and red arrows indicate the wave, and the yellow arrow points out coronal loops L1. The white arrow shows the dimmings around the end of L1, and the solid lines (S1-S3) starting at triangles are used to derive the speeds that were shown in time-distance plots (bottom panels). The dotted lines with asterisks are used to calculate the wave speeds. The green arrow indicate the jet. (An animation of W1 in AIA 211~{\AA} and EUVI-A 195~{\AA} is available online, and the animated sequence runs from 04:30 to 05:40 UT on 2010 August 18.)
\label{f3}}
\end{figure}

\clearpage

\begin{figure}
\epsscale{0.95}
\plotone{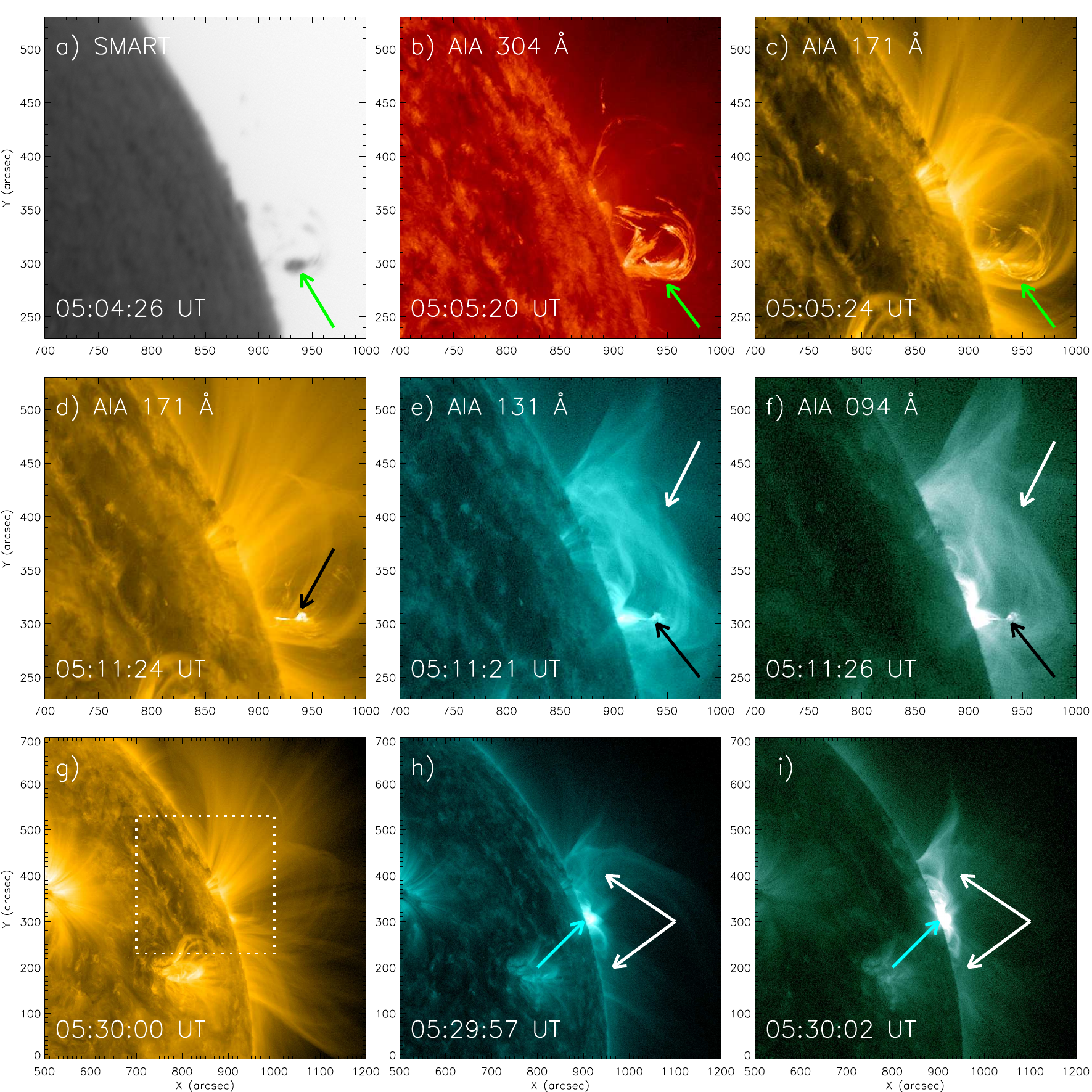}
\caption{The eruption of the newly-formed long filament in H$\alpha$ negative filtergram and AIA 304, 171, 131, and 94~{\AA}. The green arrows indicate the newly-formed filament, and the black arrows indicate the current sheet. The cyan arrows indicate post-flare arcades, and the white arrows indicate the hot channel. The dotted box in panel (g) represents the field of view (FOV) of panels (a)-(f).
\label{f4}}
\end{figure}

\clearpage

\begin{figure}
\epsscale{0.8}
\plotone{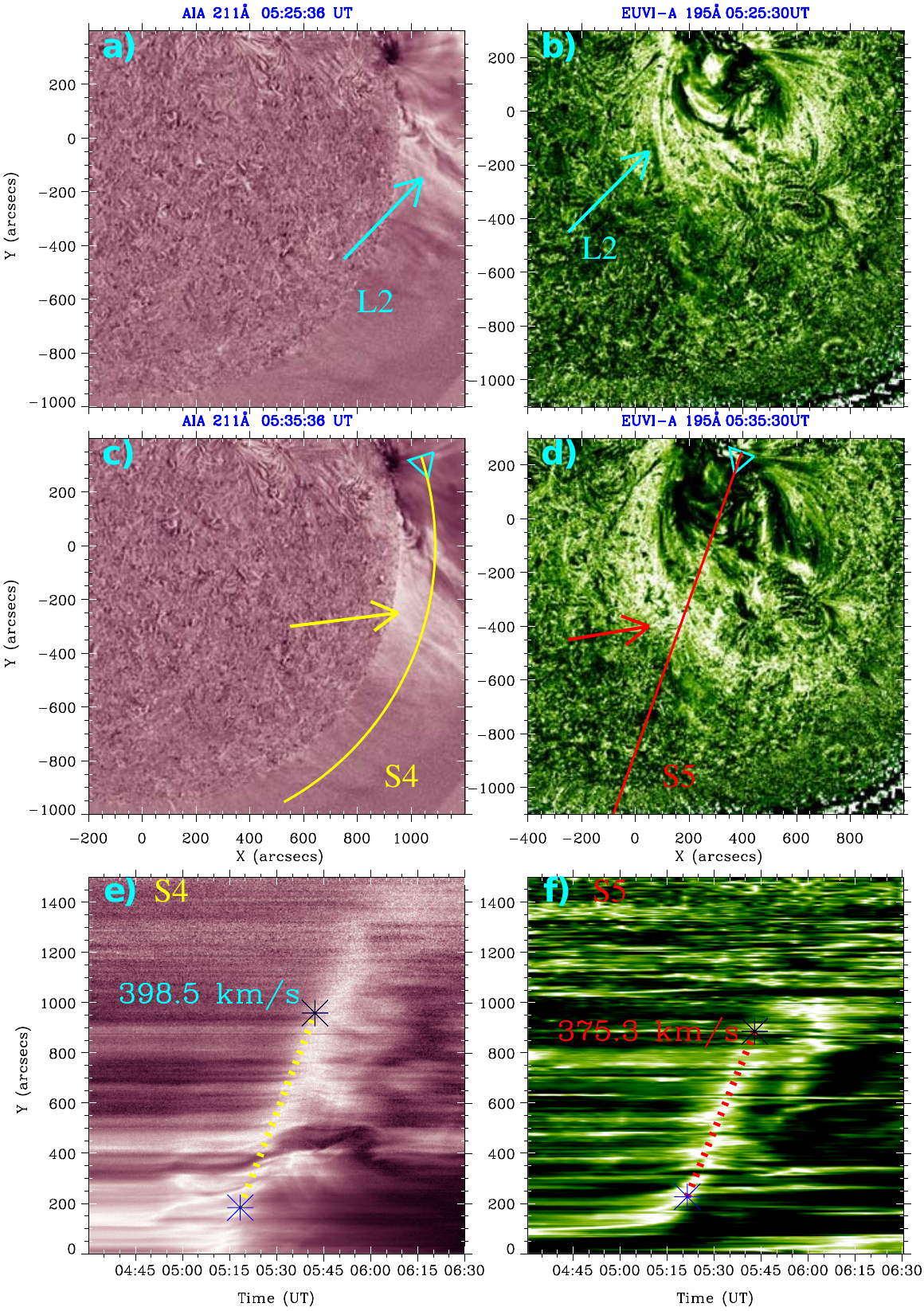}
\caption{W2 in intensity-normalized images between 04:30 and 06:30 UT in AIA 211~{\AA} (left panels) and EUVI-A 195~{\AA} (right panels). The cyan arrows point out the coronal loops L2, and yellow and red arrows indicate W2, and the solid lines (S4 and S5) starting at triangles are used to derive the wave speeds that were shown in time-distance plots (bottom panels). The dotted lines with asterisks are used to calculate the wave speeds. (An animation of W2 in AIA 211~{\AA} and EUVI-A 195~{\AA} is available online, and the animated sequence runs from 05:00 to 06:00 UT on 2010 August 18.)
\label{f5}}
\end{figure}

\clearpage

\begin{figure}
\epsscale{0.9}
\plotone{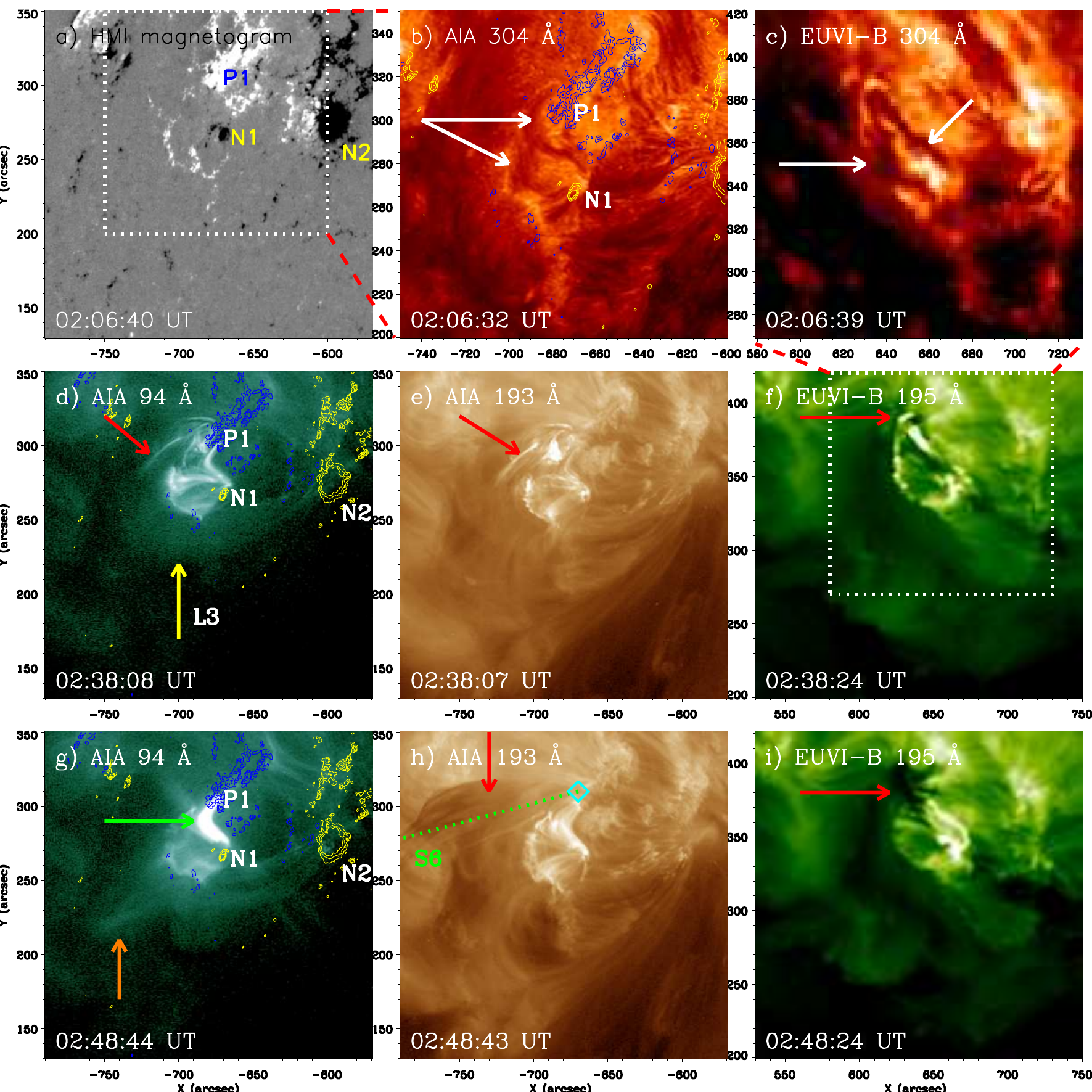}
\caption{The filament eruption in the HMI magnetogram, AIA 304, 193, and 94~{\AA}, and EUVI-B 304 and 195~{\AA}. The white arrows indicate the filament, and P1, N1, and N2 represent associated magnetic polarities. The red arrows indicate the erupting filament, and the green arrow points out the flare loops. The yellow and orange arrows indicate coronal loops L3 and the cusp structure, respectively. The dotted boxes represents the FOV of panel (b) and (c), respectively. The yellow and blue contours superposed on panel (b) represent the positive and negative magnetic polarities with levels of 200, 400, 600 Gauss. The dotted line (S6) starting at a diamond is used to derive the filament speed that is shown in the time-distance plot of Figure 7.
\label{f6}}
\end{figure}

\clearpage

\begin{figure}
\epsscale{0.95}
\plotone{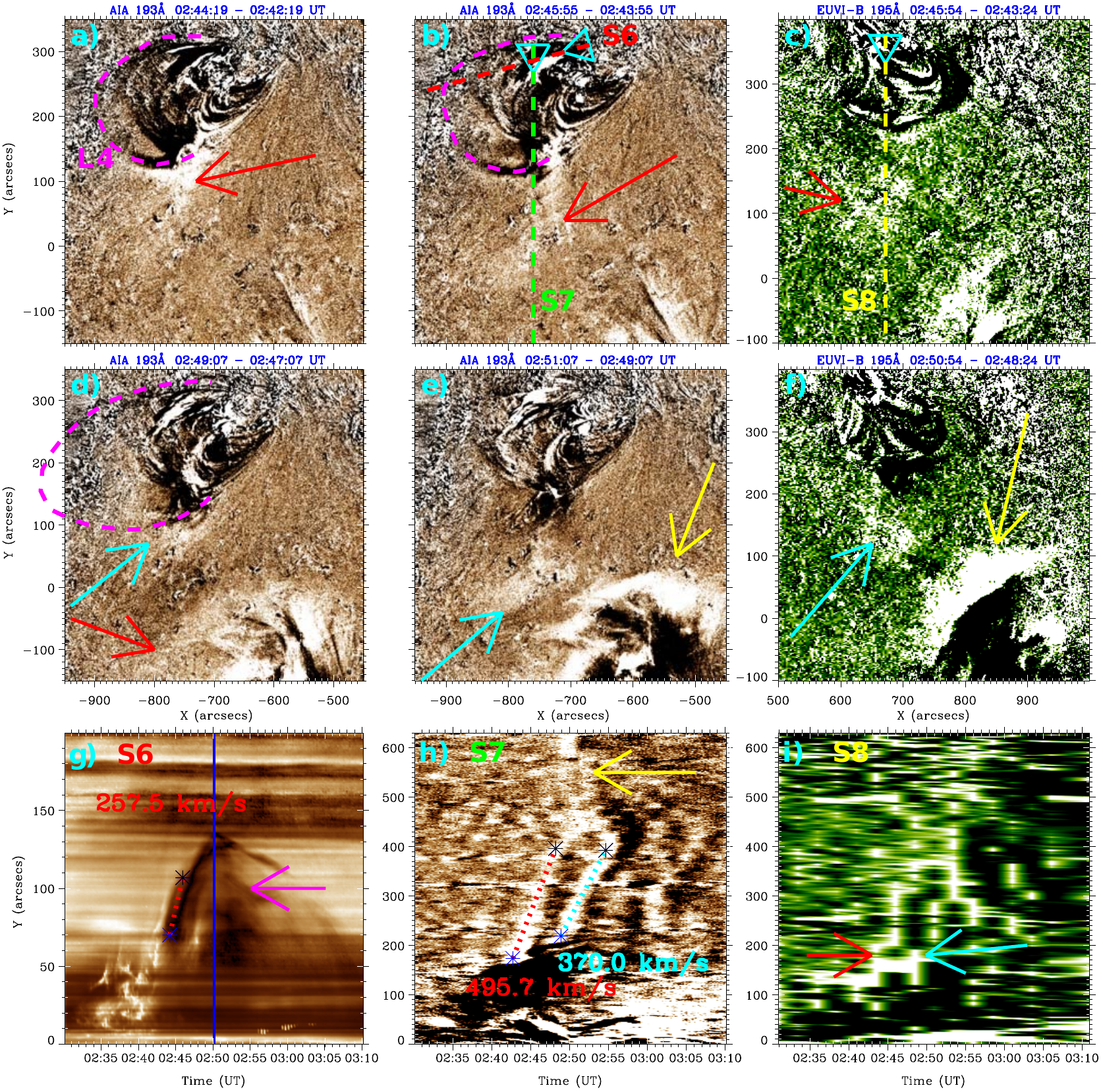}
\caption{W3 and W4 in running difference images in AIA 193~{\AA} and EUVI-B 195~{\AA}. The red and cyan arrows indicate W3 and W4, and the yellow arrow show the wave that interacted with W3 and W4. The blue vertical line shows the beginning of the falling back of the filament, and the pink arrow shows the falling material. The dashed lines (S6-S8) starting at triangles are used to derive the wave speeds that were shown in time-distance plots (bottom panels), and the dotted lines with asterisks are used to calculate the speeds. (An animation of W3-W4 in AIA 193~{\AA} and EUVI-B 195~{\AA} is available online, and the animated sequence runs from 02:35 to 03:00 UT on 2011 June 1.)
\label{f7}}
\end{figure}

\clearpage

\begin{figure}
\epsscale{0.9}
\plotone{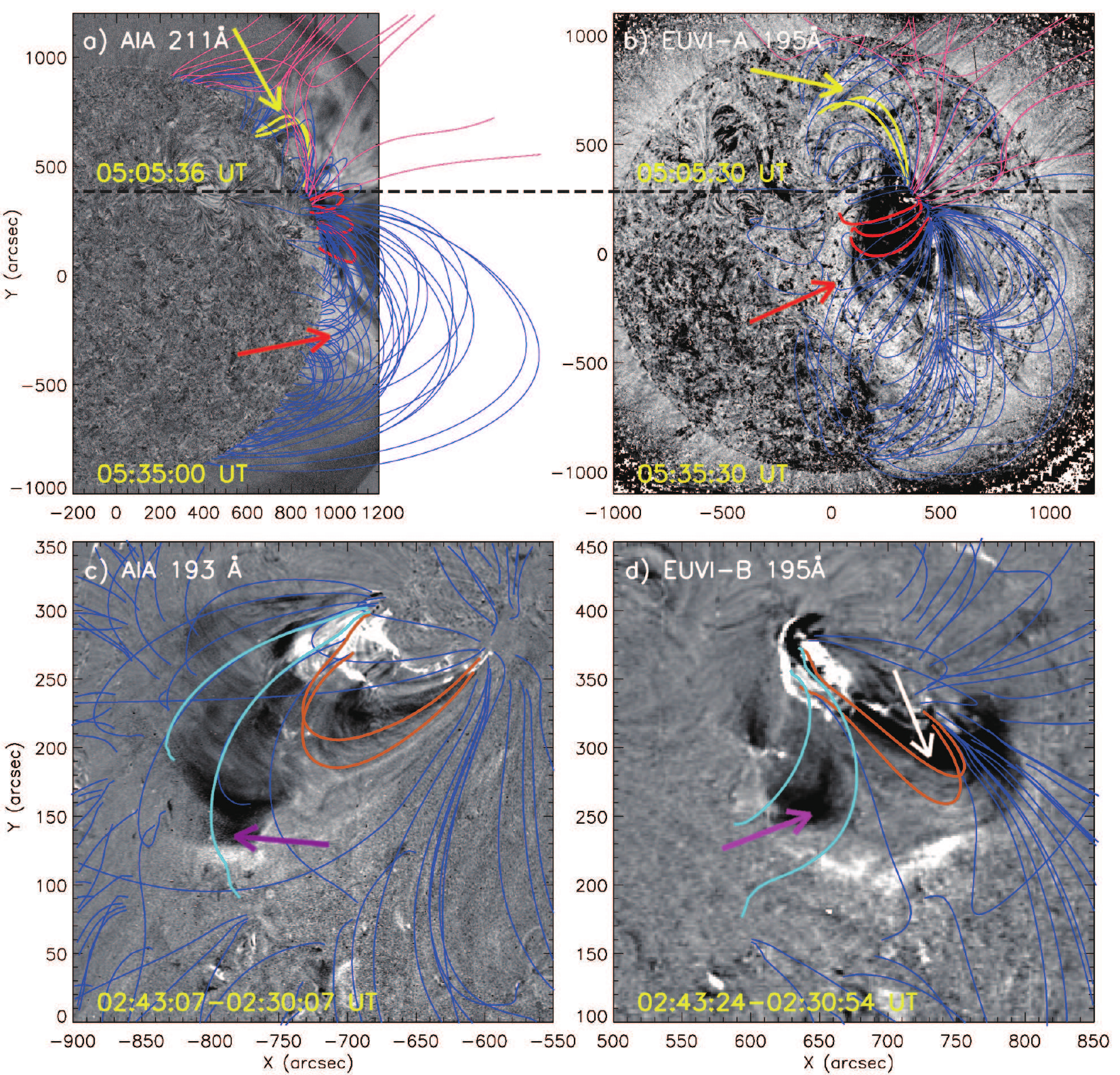}
\caption{The extrapolated PFSS field lines superimposed on the intensity-normalised images in AIA 211~{\AA} and EUVI-A 195~{\AA} (top panels) and the difference images in AIA 193~{\AA} and EUVI-B 195~{\AA} (bottom panels). The blue and pink lines represent the coronal closed loops and open field lines, respectively. The yellow, green, cyan, and orange lines are the vital loops (L1-L4) for the generation of TEWs. The yellow and red arrows show the fraternal TEWs, and the white and purple arrows show the coronal dimmings associated with identical TEWs.
\label{f8}}
\end{figure}

\clearpage

\begin{figure}
\epsscale{0.9}
\plotone{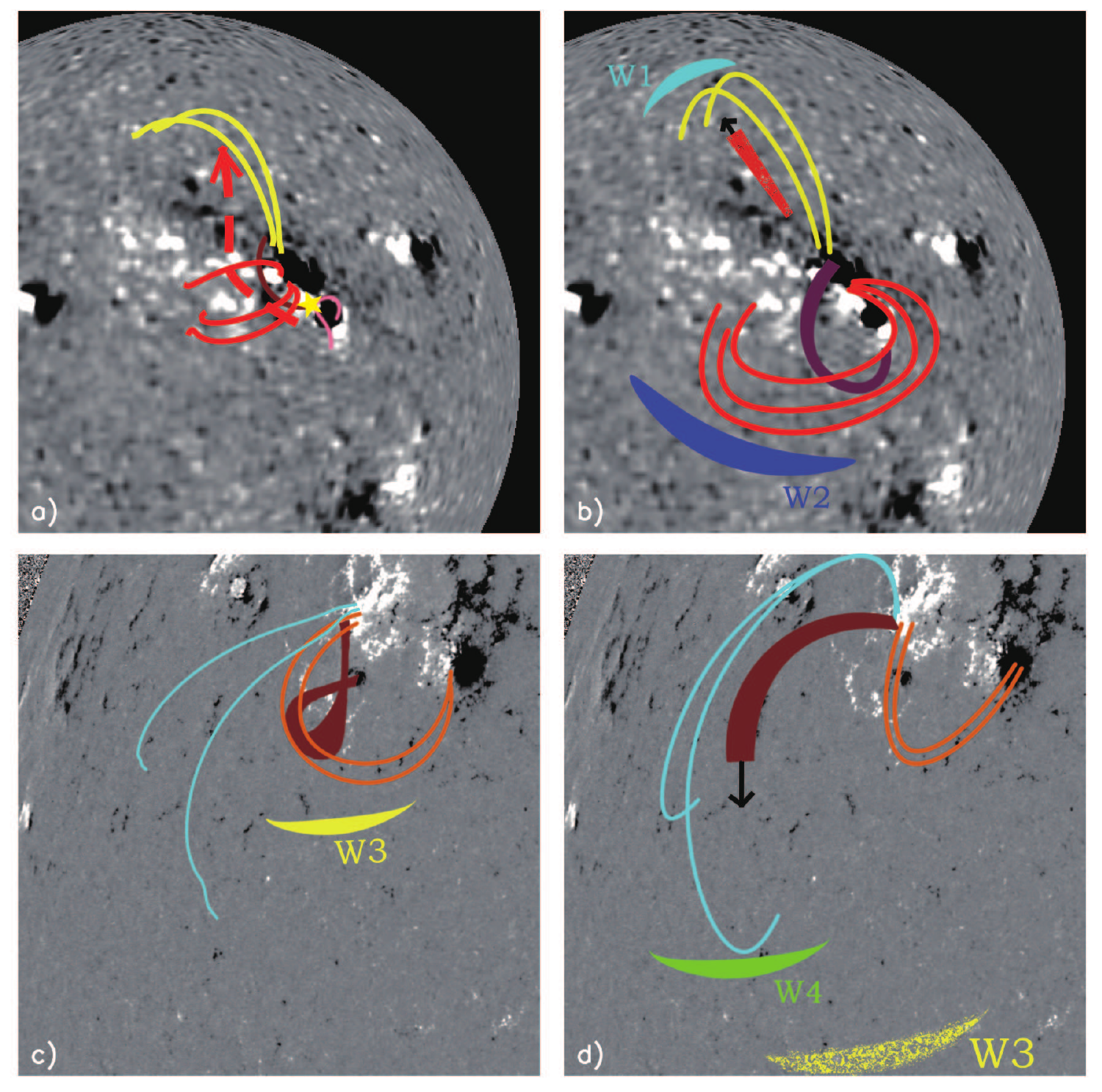}
\caption{The sketch of fraternal (top panels) and identical (bottom panels) TEWs based on the PFSS extrapolated field lines and HMI magnetograms. The yellow, green, cyan, and orange lines are the vital loops for the generation of twin waves, and the cyan, blue, yellow, and green shades represent the wavefronts of W1-W4. The yellow star indicates the magnetic reconnection between two small filaments (brown and pink lines), and the red dashed arrow represents the trajectory the jet. The thick lines show the erupting filaments or jets. The black arrows indicate the moving direction of the erupting jet and filament.
\label{f9}}
\end{figure}

\clearpage

\appendix
\section{Fraternal Twin EUV waves}
\begin{figure}
\epsscale{0.9}
\plotone{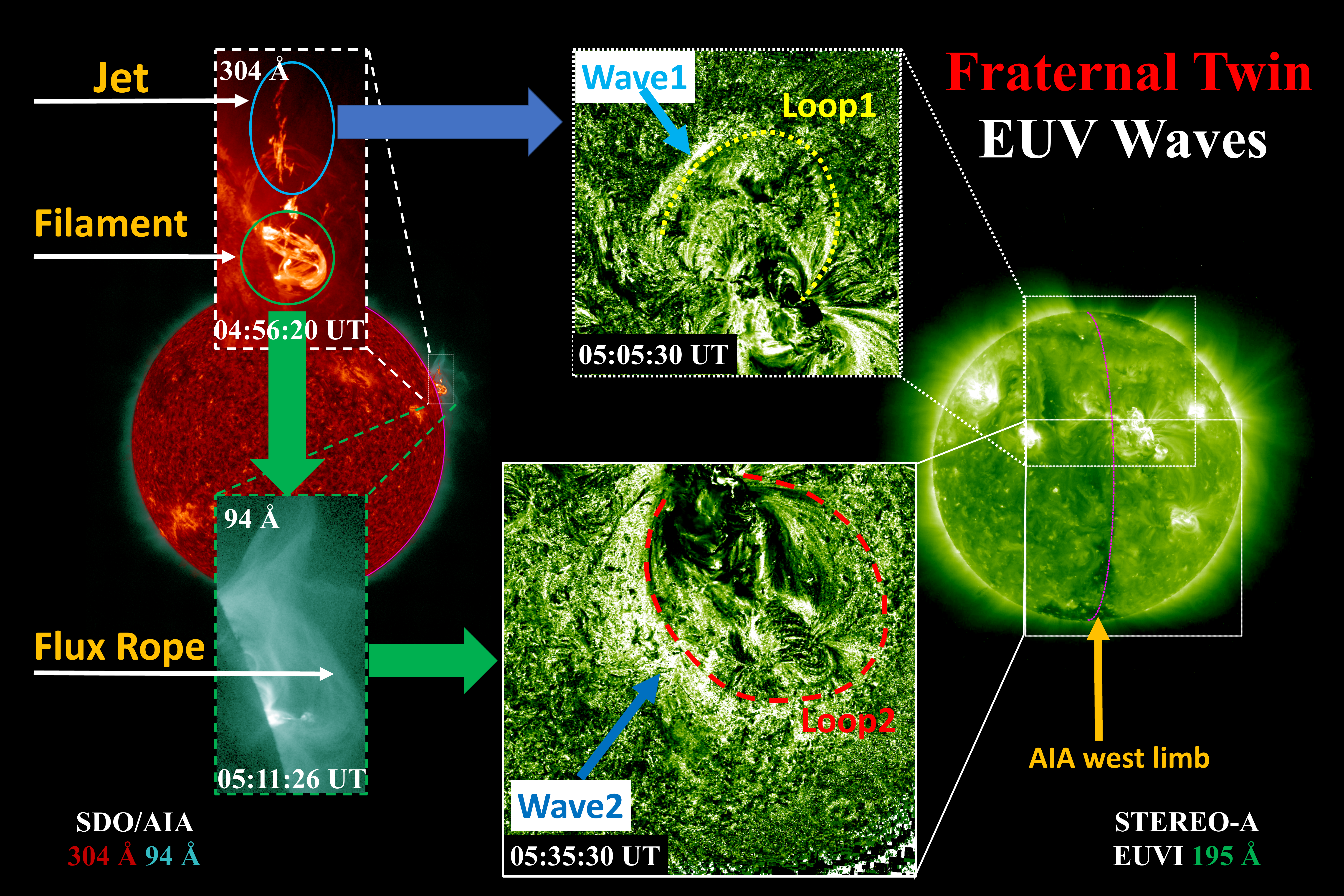}
\caption{Graphical abstract of Fraternal Twin EUV waves. Two waves are shown clearly in the running difference images in EUVI-A 195~{\AA} (middle panels), and the FOV of middle panels are indicated by the rectangles in the EUVI-A 195~{\AA} image of the right panel. The source region is represented by the dashed panel in the composite image of AIA 304 and 94~{\AA} (the left background panel), and is magnified to display the jet, filament, and flux rope in left foreground panels. The wide blue and green arrows point out the temporal evolution of waves.
\label{f10}}
\end{figure}

\clearpage  

\section{Identical Twin EUV waves}
 \begin{figure}
\epsscale{0.9}
\plotone{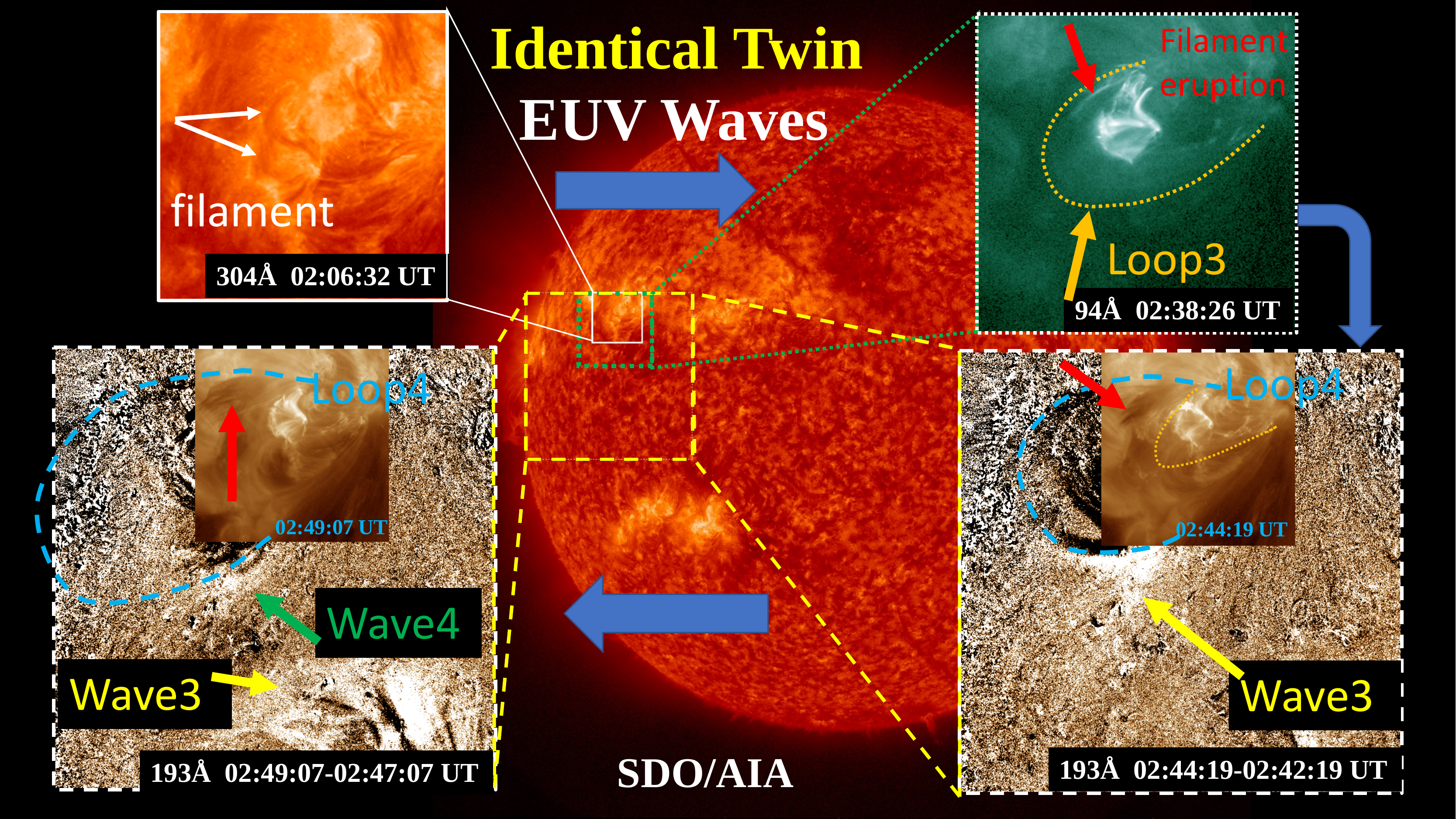}
\caption{Graphical abstract of Identical Twin EUV waves. Two waves are shown clearly in the running difference images in AIA 193~{\AA} (bottom panels), and the associated filament and the eruption are shown in images in AIA 304 and 94~{\AA} (top panels). Their FOV are indicated by the rectangles in the background image in AIA 304~{\AA}. The wide blue arrows point out the temporal evolution of waves.
\label{f11}}
\end{figure}

\clearpage 

\end{document}